%% file: main.tex
\documentclass{article}

\PassOptionsToPackage{numbers, sort&compress}{natbib}
\usepackage[dblblindworkshop, final]{neurips_2026}
\workshoptitle{Real-Time Conversational Agents: Toward Natural Multimodal Interaction}

\usepackage[utf8]{inputenc}
\usepackage[T1]{fontenc}
\usepackage{hyperref}
\usepackage{url}
\usepackage{booktabs}
\usepackage{amsfonts}
\usepackage{xcolor}
\usepackage{graphicx}
\usepackage{enumitem}

\title{PAIR: Perceptual Affective Inference and Regulation\\in a Real-Time Multimodal Conversational Agent}

\author{%
  Kexin Quan$^{1}$\thanks{Corresponding author (kq4@illinois.edu).} \enspace
  Zijian Ding$^{2}$ \enspace
  Jiaye Yong$^{1}$ \enspace
  Qinshi Zhang$^{3}$ \enspace
  Dong Wang$^{1}$ \enspace
  Jessie Chin$^{1}$ \\[0.6ex]
  \normalfont
  $^{1}$School of Information Sciences, University of Illinois at Urbana-Champaign \\
  $^{2}$School of Computing, KAIST \\
  $^{3}$Department of Computer Science, Florida State University
}

\begin{document}

\maketitle

\begin{abstract}
Sustained emotional support requires generative agents to connect momentary
emotion inference and regulation with continuity across encounters.
We present PAIR (Perceptual Affective Inference and Regulation), a real-time multimodal agent that reconstructs how an event is appraised into an emotional state.
Appraisal scaffolds produce a valence-arousal-dominance estimate and select regulation guidance, delivered through conversation with coordinated speech, color, and avatar cues.
Rolling memory carries context across sessions, and the scaffold re-runs after guidance.
In a 14-day deployment with 19 participants, 1{,}093 sessions paired initial and post-guidance estimates with unanchored self-reports.
Initial valence reached MAE 1.20 on the 9-point SAM scale ($r{=}.68$), dominance reached MAE 1.30, and arousal showed weak agreement even after coarsening.
Self-reported emotional change varied with initial state, with the largest valence increases in sessions that began at negative valence.
Perceived understanding was associated with greater valence increase and showed little correspondence with numerical prediction error.
Over two weeks, helpfulness increased while input shortened; interviews traced personalization and companionship to relevant recall, context updates, and familiar dialogue.
These findings connect inference accuracy to conversational and temporal patterns of support through per-event, first-person evaluation.
\end{abstract}

\input{sections/1-introduction}
\input{sections/2-relatedWork}

\input{sections/3-system}
\input{sections/4-evaluation}
\input{sections/5-results}
\input{sections/6-discussion}

\clearpage
\bibliography{references}
\bibliographystyle{plainnat}

\section*{LLM usage statement}
LLMs are core components of the system under study; the models serving emotional reasoning, routing, memory summarization, and speech synthesis are named in Sec.~\ref{sec:system}. In preparing this submission, LLMs were additionally used to assist with manuscript drafting and revision, analysis scripting, and figure generation. The analysis also used manually reviewed LLM-assisted classification of adoption-probe responses.

\clearpage
\appendix
\input{sections/appendix}

\end{document}

%% file: sections/1-introduction.tex
\section{Introduction}
\label{sec:intro}

Real-time conversational agents are increasingly deployed in emotionally charged settings such as companionship, coaching, and mental health support~\cite{nepal2024mindscape, fitzpatrick2017woebot, maples2024loneliness}. Support requires understanding what the user feels and why, selecting a response, and retaining relevant context. Few systems verify the interpretation behind that response against the user's own experience.

Connecting emotion recognition to supportive responses requires reasoning about how an event produces emotion and which response fits. Multimodal models recognize emotion from acoustic, visual, and textual signals~\cite{cheng2024emotionllama}, benchmarks probe the emotion competence of LLMs~\cite{huang2024emotionbench, carebench2025}, and empathetic generators produce supportive replies across modalities~\cite{fei2024empathyear, avamerg2025}. Offline benchmarks, third-party judgments, and simulated users~\cite{huang2024emotionbench, tak2023gpt, ye2026emoharbor} characterize model responses; everyday support additionally requires first-person evaluation across conversations.

We present PAIR (Perceptual Affective Inference and Regulation), a real-time multimodal conversational agent that reconstructs how emotion forms in response to a specific event. A lightweight router dispatches each narration to one of two appraisal-theoretic reasoning scaffolds~\cite{lazarus1991emotion, scherer2001appraisal}, selected through formative human evaluation. Both produce an estimate over valence, arousal, and dominance (PAD)~\cite{mehrabian1996pleasure}, together with appraised intensity and controllability. These attributes select among four regulation families from Gross's process model~\cite{gross2015emotion}, and the selected strategy unfolds through a multi-turn conversation. A shared regulation target coordinates speech prosody, ambient color, and avatar animation.

A rolling memory carries prediction discrepancies, corrections, and guidance outcomes into later sessions without gradient updates. After each conversation, the same scaffold re-runs on the transcript to estimate the post-guidance state, exposing both initial interpretation and subsequent assessment to evaluation.

We deployed PAIR for 14 days with 19 participants, yielding 1{,}093 completed sessions. Each initial and post-guidance estimate was paired with the participant's own Self-Assessment Manikin (SAM) rating~\cite{bradley1994measuring}, with numerical estimates withheld to avoid anchoring. Session logs, feedback, and exit interviews connect these event-level comparisons to three research questions.

\begin{enumerate}[leftmargin=*, itemsep=1pt, topsep=2pt]
    \item \textbf{RQ1 (Emotion Inference):} How accurately does text-based appraisal inference estimate users' valence, arousal, and dominance from everyday emotional events?
    \item \textbf{RQ2 (Regulation Outcomes):} How do initial emotional state, regulation strategy, and guidance fit relate to momentary regulation and guidance uptake?
    \item \textbf{RQ3 (Long-Term Companion):} How do conversational familiarity, cross-session memory, and accumulated personal context shape personalization and companionship across a 14-day deployment?
\end{enumerate}

Our contributions are \textbf{1)} an integrated architecture linking appraisal reasoning, contextual regulation, and multimodal delivery; \textbf{2)} an in-situ protocol pairing initial and post-guidance estimates with first-person reports across everyday encounters; and \textbf{3)} findings on dimension-specific inference accuracy, state-dependent emotional change, and the conversational and memory processes shaping support over two weeks.

%% file: sections/2-relatedWork.tex
\section{Related Work}
\label{sec:related}

\begin{figure}[t]
  \centering
  \includegraphics[width=\linewidth]{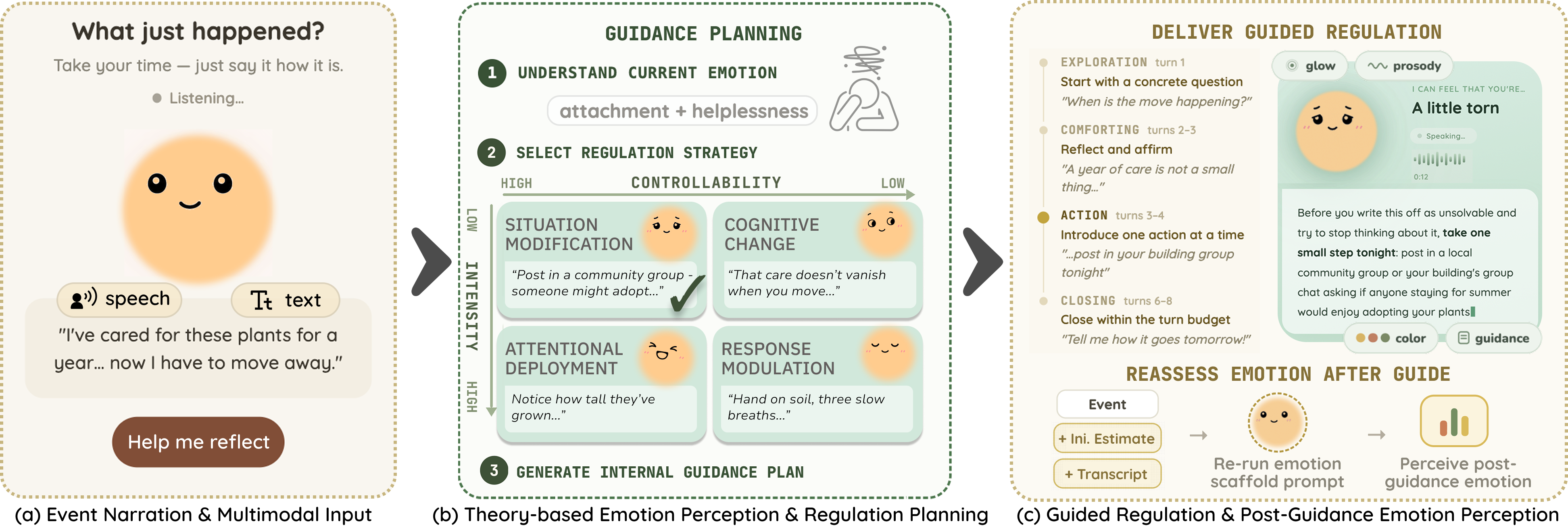}
  \caption{PAIR pipeline. Event narration enters an appraisal scaffold that estimates emotional state and selects a regulation strategy. A guided conversation unfolds the plan with coordinated multimodal delivery, followed by a second estimate using the conversation transcript.}
  \label{fig:pipeline}
\end{figure}

\paragraph{Emotion reasoning in conversational agents.}
Emotion reasoning research connects affective judgments to multimodal context, explicit appraisal, and personal information. Emotion-LLaMA integrates multimodal recognition and reasoning~\cite{cheng2024emotionllama}, while EmotionBench, EmoBench, and CAREBench examine how model judgments align with human affective assessments~\cite{huang2024emotionbench, sabour2024emobench, carebench2025}. Explicit appraisal dimensions ground generated cognitive reappraisals~\cite{zhan2024reappraisal}. PersonaAgent further personalizes agents through test-time persona adaptation~\cite{zhang2025personaagent}. PAIR uses appraisal output for strategy selection and accumulated feedback for subsequent context.

\paragraph{Multimodal generation for conversational agents.}
Empathetic generation extends emotional expression across text, speech, and avatars. EmpathyEar generates avatar, speech, and text responses~\cite{fei2024empathyear}; AvaMERG studies avatar-based empathetic generation~\cite{avamerg2025}; and WeSCon provides word-level emotional control of speech~\cite{wescon2025}. PAIR coordinates these expressive channels through the emotional direction targeted by regulation.

\paragraph{Evaluation of live systems.}
Participant annotations and field deployments extend evaluation to experienced support within and across conversations. AttuneBench collects participant annotations across genuine dialogues~\cite{attunebench2026}, and HEART compares human and LLM responses in support conversations~\cite{heart2026}. Field deployments include a seven-user health coach~\cite{ozolcer2025health}, MindScape's eight-week journaling study with 20 students~\cite{nepal2024mindscape}, and a four-week study of affective chatbot use with 981 participants~\cite{mitopenai2025}. Our evaluation connects event-level estimates and self-reports to guidance, personalization, and companionship across encounters.

%% file: sections/3-system.tex
\section{System Architecture}
\label{sec:system}

PAIR links emotion inference to regulation guidance and multimodal delivery (Figure~\ref{fig:pipeline}). Claude Opus 4.8 performs emotion reasoning, while GPT-5-mini handles routing and memory summarization. Both typed narration and browser-transcribed speech are processed as text for inference.

\subsection{The Emotion Inference Path}
\label{sec:inference}

\begin{figure}[t]
  \centering
  \includegraphics[width=\linewidth]{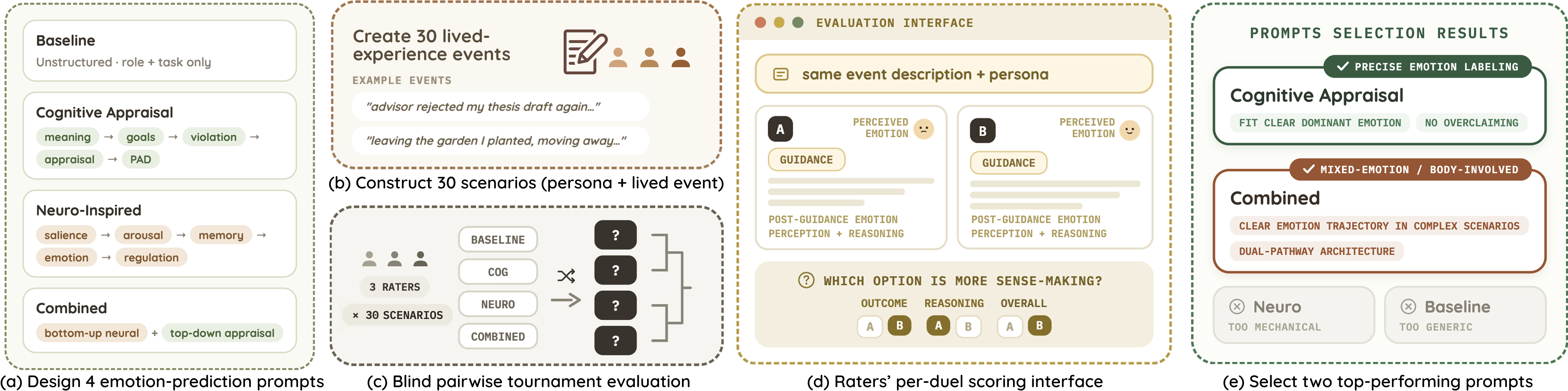}
  \caption{Prompt evaluation pipeline. (a) Four scaffolds and (b) 30 persona--event scenarios enter (c) blind pairwise tournaments with (d) three rating dimensions, (e) yielding COG and COMBINED for routing.}
  \label{fig:prompteval}
\end{figure}

\textbf{Scaffold design and event-adapted routing.}\label{sec:router} A blind comparison of four emotion-prediction prompts identified complementary scaffolds for single- and multi-component events (Figure~\ref{fig:prompteval}). The tournament used 30 persona--event scenarios, three trained raters, and 1{,}080 pairwise judgments (Appendix~\ref{app:scaffolds}). The cognitive-appraisal scaffold (\textbf{COG}) follows a five-stage appraisal chain \cite{lazarus1991emotion, scherer2001appraisal}. \textbf{COMBINED} extends this scaffold with bottom-up neural reasoning for events involving interacting bodily and cognitive components \cite{ledoux2000emotion, pessoa2008relationship}. The raters compared anonymized outputs from the same model on outcome plausibility, reasoning quality, and overall preference. Separately, two researchers coded each scenario as single- or multi-component without consulting the rankings. COG and COMBINED each won 15 scenarios, and event structure matched the preferred scaffold in 26 of 30 scenarios (87\%). The router therefore directs single-component events to COG and multi-component events to COMBINED.

\textbf{Prediction output and per-user grounding.}\label{sec:prediction} Each prediction is grounded in the user's onboarding persona, a calibration event paired with self-reported SAM ratings, and rolling memory from prior encounters. Both scaffolds output continuous PAD estimates, predicted SAM ratings on the user's 1--9 scale, emotion labels and intensity, controllability, and temporal phase. After each evening session, a summarizer updates this memory by compressing recurring themes, prediction discrepancies, recent corrections, regulation outcomes, and the previous four days of context into at most 150 words (Appendix~\ref{app:memory}).

\subsection{The Regulation Path}
\label{sec:regulation}

\textbf{Strategy selection and guided conversation.}\label{sec:strategy} The appraisal call selects among four regulation strategies: situation modification, attentional deployment, cognitive change, and response modulation~\cite{gross2015emotion}. Strategy selection considers emotional intensity, which guides the balance between immediate relief and reappraisal~\cite{sheppes2011emotion}, and controllability, which guides whether to modify the situation or its interpretation~\cite{troy2013cognitive, haines2016wisdom}. Temporal phase, touchpoint, and personal context further refine the selection (Appendix~\ref{app:strategy}). The system then generates an internal plan specifying the regulatory move and a concrete first step. The conversation follows an Exploration--Comforting--Action (ECA) arc~\cite{liu2021towards}, first eliciting context, then acknowledging the participant's emotion, and finally introducing guidance within a ten-turn budget.

\textbf{Post-guidance estimation.}\label{sec:post} After conversation, the selected appraisal scaffold runs again using the original narration, initial estimate, and conversation transcript to estimate the participant's post-guidance state. This estimate is evaluated against the participant's post-guidance SAM report.

\subsection{Coordinated Multimodal Delivery}
\label{sec:multimodal}

A shared regulation target coordinates speech prosody, ambient color, and avatar pulse to convey the emotional direction of guidance. GPT-4o-mini-TTS renders speech using directions for pace, energy, pitch, and pauses; ambient hue follows emotion--color associations, with brightness and saturation varying with target valence and arousal; and an avatar pulse slows for lower-arousal targets. A lightweight expression selector also reads the accumulating transcript and updates the avatar's face during interaction. Users can toggle speech output.

%% file: sections/4-evaluation.tex
\section{Evaluation: 14-Day Field Deployment}
\label{sec:evaluation}

\paragraph{Participants and protocol.}
We recruited 24 U.S. adults through university and social-media channels; five withdrew and 19 completed the study (13 women, 6 men; age $M{=}25.1$, $SD{=}5.1$, range 18--33). High AI familiarity was reported by 68\%, and prior emotional-support use by 42\%. Eligible participants were not receiving psychological or psychiatric treatment. The study received university IRB approval, and compensation was up to \$80, prorated by completion. Consent explained the system's non-clinical role and provided mental-health resources.

Each session paired narration and guided conversation with before/after SAM reports, four feedback items, and an understanding check. Day-0 onboarding covered consent, surveys, a persona interview, and calibration. Over 14 days, participants completed morning check-ins, self-initiated just-in-time logs, and evening reflections. Numerical estimates were withheld; understanding ratings at or below 4 prompted corrections. Alternating evenings included a rotating probe. All completers answered a 13-item exit questionnaire and a semi-structured interview.

\paragraph{Measures and analysis.}
We evaluated inference, regulation, and cross-session experience by combining paired SAM reports, logs, and feedback from 1{,}093 completed post-onboarding sessions with interviews from all 19 completers. Accuracy measures were MAE, Pearson $r$, signed error, and coarse-grained Cohen's $\kappa$. Initial valence was also compared with an unscaffolded same-model baseline and a five-fold cross-validated NRC-VAD lexicon estimator~\cite{mohammad2018vad}. SAM change was summarized overall and by initial emotional state. Wilcoxon tests used participant mean changes, with Holm correction across the three dimensions; strategy comparisons used participant-level Friedman tests with permutation inference. Understanding correlations used whole-participant bootstrap intervals, and correction contrasts used paired participant means. Mixed-effects models estimated day and week contrasts, with false-discovery-rate (FDR) correction across ten measures per model. Strategy-feedback models controlled participant and participant-day clustering. Two researchers independently coded interviews and corrections and reconciled differences; 44 adoption-probe classifications received manual review after LLM-assisted coding.

%% file: sections/5-results.tex
\section{Results}

\subsection{Overview of Deployment and Engagement}
Participants used PAIR for brief conversations, most often through self-initiated just-in-time logs and typed narration. They completed 1{,}093 sessions containing 10{,}953 conversation turns, averaging 57.5 sessions each ($SD{=}11.8$, range 35--78). Just-in-time logs accounted for 601 sessions (55.0\%), morning check-ins for 258 (23.6\%), and evening reflections for 234 (21.4\%). Guided conversations averaged 10.02 user and companion turns ($SD{=}2.80$), and the median session lasted under three minutes. Narration was typed in 872 sessions (79.8\%) and spoken in 221 (20.2\%); five participants used voice in more than 40\% of sessions.

\subsection{RQ1: Appraisal Inference Recovers Valence More Closely Than Arousal}

\begin{table}[t]
\centering\small
\begin{tabular}{@{}llrrr@{}}
\toprule
Target & Dimension / estimator & MAE & Pearson $r$ & Signed error \\
\midrule
Initial & Valence   & 1.196 & .681 & $+.213$ \\
 & Dominance & 1.299 & .538 & $-.061$ \\
 & Arousal   & 1.651 & .236 & $-.129$ \\
\midrule
Post-guidance & Valence   & 1.115 & .614 & $-.128$ \\
 & Dominance & 1.189 & .495 & $-.393$ \\
 & Arousal   & 1.963 & .175 & $-1.097$ \\
\midrule
Valence comparison & NRC-VAD lexicon, 5-fold CV & 1.679 & .318 & --- \\
 & LLM without scaffold & 1.334 & .640 & --- \\
 & Full appraisal scaffold & 1.196 & .681 & --- \\
\bottomrule
\end{tabular}
\caption{Accuracy on 1--9 SAM scales for 1{,}093 sessions, followed by a valence comparison rerun on all narrations after deployment. Both LLM conditions use Claude Opus 4.8. Signed error is prediction minus self-report.}
\label{tab:accuracy}
\end{table}

\subsubsection{Appraisal-Guided Estimates Show Stronger Valence and Dominance Agreement}

Initial estimates tracked self-reported valence and dominance more closely than arousal, while participants valued interpretations that connected feelings to their causes. Valence reached $r{=}.681$ (MAE 1.196) and dominance $r{=}.538$ (MAE 1.299; Tab.~\ref{tab:accuracy}). In the separate valence comparison, MAE fell from 1.679 for NRC-VAD to 1.334 for the unscaffolded model and 1.196 with appraisal. The largest descriptive gain came from the model's contextual reading; the scaffold additionally supplied the intensity and controllability used for regulation. Fourteen participants described cases in which the companion inferred feelings or their causes without explicit emotion words. P1 valued its ability to go \textit{``straight from my description to dissecting what was going on inside me,''} and P10 valued identifying \textit{``what was responsible''} for an emotion she could not name. Exit ratings were highest for understanding feelings in the moment ($M{=}5.63$, $SD{=}.68$), followed by post-conversation understanding (5.37, .90) and emotional nuance (5.21, .79).

\subsubsection{Text-Only Estimates Show Weak Arousal Agreement and Intensity Mismatches}

PAIR's arousal estimates showed weak agreement with self-report, and participants described emotional direction being recognized while intensity was misjudged. Speech input was transcribed before inference, leaving prosody unavailable alongside facial and bodily cues. Its arousal estimates therefore reflected what participants said in the transcript without information from how they said it. Arousal correlated weakly with self-report ($r{=}.236$) and was underestimated after guidance by 1.097 points on average. Coarsening to low/high ratings did little to resolve the mismatch ($\kappa{=}.154$; Appendix~\ref{app:coarse}); no tested arousal cut point exceeded $.21$, and NRC-VAD likewise showed little correspondence with self-report ($r{=}.016$). Eight participants described intensity mismatches. P12 found her emotions taken \textit{``way too extreme,''} while P3 distinguished venting from actually being angry. Participants sometimes compensated by softening wording (P12) or explicitly stating intensity (P18), linking this input boundary to how they communicated with the companion.

\subsection{RQ2: Initial Emotional State and Guidance Fit Relate to Momentary Regulation}

\subsubsection{Negative Initial Emotional States Show the Largest Valence Changes}

\begin{figure}[t]
  \centering
  \includegraphics[width=\linewidth]{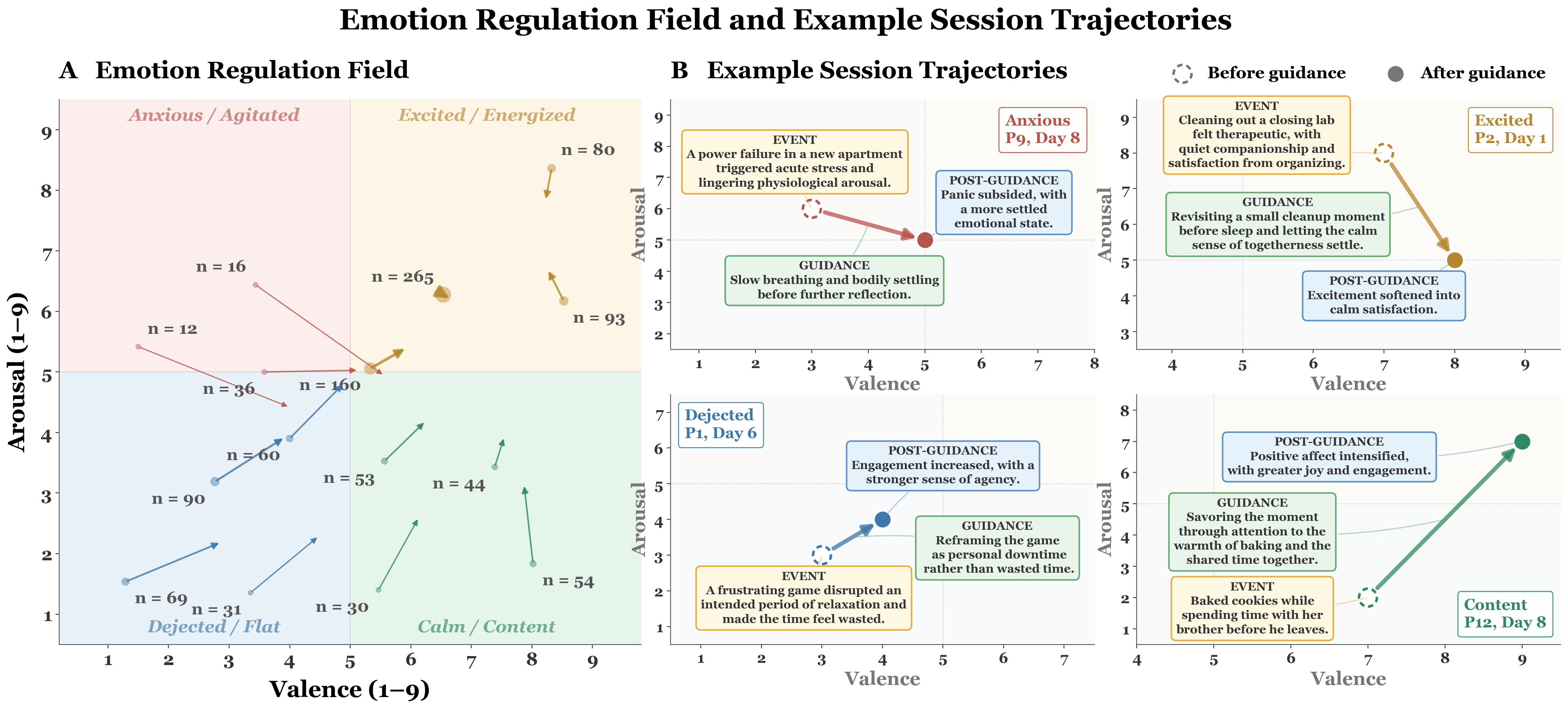}
  \caption{State-dependent affective movement. (A) Arrows connect clustered mean self-reports before and after guidance. Anxious and dejected initial states generally move toward higher valence, with opposite arousal directions. (B) One session from each initial-state quadrant links narration and guidance to the reported change. Quadrants split both dimensions at 5.}
  \label{fig:regulation_field}
\end{figure}

Negative initial emotional states showed the largest valence increases, with arousal decreasing in anxious sessions and increasing in dejected sessions. Across sessions, mean pre-post changes were .525 in valence, .469 in dominance, and .334 in arousal (all participant-level Wilcoxon $p_{\mathrm{Holm}}{<}.001$; Tab.~\ref{tab:delta}). Low-valence sessions ($V{\leq}4$) increased by 1.344 valence points, compared with $-.108$ for high-valence initial states ($V{\geq}7$). Anxious initial states increased by 1.875 valence points while arousal fell by .547; dejected initial states increased by 1.208 and .780 points respectively (Figure~\ref{fig:regulation_field}). Excited and calm initial states showed smaller valence increases. Participants located benefit in states they wanted to change. P10 entered feeling drained and left feeling better, and P16 valued support when exhausted. Others entered with little distress to resolve. Narration also mattered, as P12 described writing a problem down as making it feel smaller.

\subsubsection{Contextual Exploration and Acknowledgment Shape Perceived Understanding}

Participants described feeling understood when the conversation clarified their situation and acknowledged their emotion before offering advice. Fourteen participants valued clarification and follow-up questions. P3 reported a good grasp after \textit{``going back and forth three, four times,''} and P16 described a clarification question followed by an attempt to empathize. These exchanges gave participants room to refine the interpretation before guidance. Questions could also miss the issue, as P8 found they were not on \textit{``the right track.''} Per-session understanding averaged 5.45 of 7 ($SD{=}.81$) and showed little correspondence with numerical prediction error. Participants never saw the numerical estimates; their judgments concerned the understanding conveyed through conversation. Corrections likewise extended beyond emotion labels to context, memory, and the desired support, with P2 correcting \textit{``the wrong suggestion.''} Perceived understanding was positively associated with valence gain ($r{=}.134$), and corrected sessions showed lower mean gains than uncorrected sessions ($-.273$ versus $+.559$; paired Wilcoxon $p{=}.005$).

\subsubsection{Contextual Strategy Selection Yielded Comparable Immediate Outcomes}

Selected strategies produced distinct regulatory actions with no detectable difference in participant-level valence gains (Friedman $\chi^2(3){=}.613$, permutation $p{=}.897$). Immediate feedback was also similar across families (Tab.~\ref{tab:strategy-feedback}; all $p_{\mathrm{FDR}}{=}.395$). Participants nevertheless recognized different moves, including starting with the easiest part of a task (situation modification, P1), reconsidering self-blame against completed work (cognitive change, P19), taking a snack break (attentional deployment, P12), and breathing before an exam (response modulation, P16). Six participants also described a recurring pool of suggestions, making contextual fit a continuing requirement across strategy families.

\subsubsection{Practical Feasibility and Support Needs Shape Guidance Uptake}

Adoption depended on available resources and the support wanted at that moment. Of 44 manually reviewed responses, 24 reported trying guidance with a positive result, five described partial or alternative enactment or mixed outcomes, and four reported actions already intended. Eight reported not trying the suggestion and three could not recall it. P7 valued being able to \textit{``take one minute to the side,''} while a walk could take more time than she had. P3 could not obtain a suggested drink at 5 a.m., and P12 would not interrupt a conversation to follow advice. Participants also distinguished reassurance (P2), motivation to act (P18), and further listening, as P7 explained that \textit{``sometimes you just want something to listen to you.''} Exit ratings favored relevant guidance and emotional coping (both $M{=}5.26$) over new perspectives (4.79) and improved emotional self-understanding (4.63).

\subsection{RQ3: Cross-Session Continuity Shapes Long-Term Personalization and Companionship}

\subsubsection{Conversational Familiarity Changes Disclosure and Support Expectations}

Participants continued using PAIR at a similar frequency while writing less and rating helpfulness more highly (Tab.~\ref{tab:trends}). Narration length and user characters per turn decreased by 3.37\% and 4.30\% per day; helpfulness increased by .029 points per day. Unadjusted helpfulness rose from 4.94 in Week 1 to 5.14 in Week 2, with a model-estimated week contrast of $+.216$ ($p_{\mathrm{FDR}}{=}.012$). Session frequency, understanding, warmth, feasibility, and willingness showed no reliable day trends after correction. Nine participants described adapting their wording or conversational approach. P5 compressed openings to \textit{``use as few words as possible''} while retaining context; P8 shortened openings to wait for follow-up questions; and P18 learned to state emotion explicitly. Shorter input also reflected withdrawal, as P16 shared less after repetitive advice. Three participants connected changing evaluations to their own goals, familiarity, or interaction skills. P5 increasingly sought practical suggestions, P8 grew more accepting of familiar responses, and P12 could not separate system improvement from improved interaction on her part. Shorter input thus encompassed both strategic adaptation and withdrawal.

\begin{table}[t]
\centering\small
\begin{tabular}{@{}llrr@{}}
\toprule
Measure & Model scale & Daily slope [95\% CI] & $p_{\mathrm{FDR}}$ \\
\midrule
Sessions per active day & log link & $-.0004$ [$-.0152$, $.0145$] & .963 \\
Conversation turns & log link & $-.0050$ [$-.0098$, $-.0002$] & .087 \\
Narration length & log characters & $-.0343$ [$-.0475$, $-.0210$] & $<.001$ \\
User input per turn & log characters & $-.0439$ [$-.0548$, $-.0330$] & $<.001$ \\
Absolute valence error & log1p SAM points & $-.0069$ [$-.0142$, $.0003$] & .087 \\
Understanding & 1--7 & $+.0121$ [$-.0004$, $.0245$] & .087 \\
Helpfulness & 1--7 & $+.0294$ [$+.0112$, $.0476$] & .005 \\
Warmth & 1--7 & $+.0174$ [$-.0002$, $.0351$] & .087 \\
Feasibility & 1--7 & $+.0134$ [$-.0045$, $.0313$] & .177 \\
Willingness to act & 1--7 & $+.0123$ [$-.0061$, $.0306$] & .211 \\
\bottomrule
\end{tabular}
\caption{Mixed-effects day trends over Days 1--14 (1{,}093 sessions; 266 observed participant-days for frequency). FDR correction covers all ten tests. Slopes are on the stated model scale.}
\label{tab:trends}
\end{table}

\subsubsection{Relevant Recall and Persistent Corrections Shape Perceived Personalization}

Participants associated personalization with relevant memories, incorporated developments, and persistent corrections. Interviews and perceived-learning ratings captured this experience of personalization, while absolute valence error showed no significant decline across days ($p_{\mathrm{FDR}}{=}.087$; Tab.~\ref{tab:trends}). Perceived learning over time averaged 5.37 of 7 ($SD{=}1.01$), while personalized emotion interpretations averaged 4.79 ($SD{=}1.40$). Ten participants described personalized responses, while fourteen also encountered generic ones. Recall became meaningful when it connected earlier experiences to later developments. P19 described the companion remembering a new after-work caf\'e routine and recognizing it as \textit{``a very good change.''} P1 acknowledged retained information but wanted more than \textit{``a fairly linear answer.''} Continuity weakened when context was missing or outdated. Twelve participants reported forgetting, and six reported temporal or contextual confusion. P10 had to \textit{``list it all over again''} after an earlier same-day conversation; P2 encountered an event from three days earlier treated as current; and P15 found occasional onboarding activities turned into recurring assumptions. Six participants reported incorporated feedback, while nine described corrections that did not persist or eventually stopped correcting. P12 observed calmer language after correcting exaggerated intensity, whereas P4 saw an immediate clarification fail to carry forward.

\subsubsection{Accumulated Records Enable Reflection and Regular Dialogue Builds Companionship}

Accumulated records provided material for reflection, while daily exchanges gave some participants a sense of companionship. Five participants valued logs for revisiting and organizing experiences, including people who did not ordinarily keep diaries. P15 described an \textit{``interactive diary,''} combining a record with responses. Nine participants valued an available listener with fewer obligations to reciprocate support or manage another person's reactions. P7 emphasized that it \textit{``will always answer you,''} while P11 valued sharing ordinary updates without burdening someone else. Four participants described the companion as a friend, often linking this feeling to daily contact. The avatar supplied a visible counterpart (P2, P18), and familiar exchanges became something to anticipate. P10 reported, \textit{``On day 15, I actually missed talking to it.''} Others described a listener or acquaintance, sought people for more serious concerns or broader involvement (P3, P4), or disclosed private material to the companion that they would hesitate to share with others (P11, P12, P19).

%% file: sections/6-discussion.tex
\section{Discussion \& Conclusion}
\label{sec:discussion}

\subsection{Ethical Considerations}
\label{sec:discussion-ethics}
Our study raises four ethical considerations for emotional support companions. \textbf{(1) Clinical boundaries.} Everyday emotional support should remain clearly separated from treatment and crisis response, given documented failures of LLMs in clinically sensitive situations~\cite{moore2025stigma}. Participants were informed of PAIR's non-clinical role and provided mental-health resources, while higher-risk use requires dedicated safety mechanisms and access to human support. \textbf{(2) Guidance quality and autonomy.} Inaccurate emotion inference, misplaced reassurance, memory errors, and sycophantic responses can undermine support or reinforce questionable judgments~\cite{cheng2026sycophantic}. Systems should preserve users' ability to reject advice, correct interpretations, and express negative emotion without pressure toward positivity. \textbf{(3) Relational attachment.} Participants' reluctance to end the study and reports of missing the companion illustrate the attachment risks documented for social chatbots~\cite{laestadius2024toohuman}. Emotional companions should support deliberate closure, avoid exclusivity or guilt-based engagement, and reinforce existing human relationships. \textbf{(4) Privacy and persistent memory.} Intimate disclosure requires clear control over what is stored and retained, particularly when companion framing can obscure organizational data practices~\cite{chiu2026privacy}. Systems should minimize stored information and support transparent data use, memory inspection, correction, and deletion. We therefore withhold study conversations and diaries from public release.

\subsection{Limitations and Future Work}
\label{sec:discussion-limitations}

Our single-group diary study deployment did not isolate the effects of guided conversation from natural emotional change, disclosure itself, regression to the mean, or other system components. Context-dependent strategy selection also prevents causal comparison across strategy families. Future studies should include unguided journaling, alternative selection policies, memory ablation, and text-only delivery as comparison conditions. Besides, self-reported affect provides a first-person reference but remains subject to individual differences in emotion awareness and scale use. The two-week deployment primarily captures early adaptation under scheduled check-ins and compensation. Longer studies with fewer required interactions are needed to assess sustained voluntary use, personalization, and attachment. Moreover, generalizability is also limited by the small, young, AI-familiar sample. Broader samples and longer-term outcome measures should test whether these findings extend across populations and persist over time.

\paragraph{Conclusion.}
\label{sec:discussion-conclusion}
Across 1{,}093 sessions with 19 participants, PAIR connected appraisal-grounded emotion inference, regulation guidance, and post-guidance estimation. The fourteen-day deployment showed dimension-specific inference accuracy and state-dependent emotional change, while participants linked their experience of support to conversational fit and updated memory. These findings frame sustained emotion regulation as a task of connecting momentary understanding with continuity across encounters. Per-event, first-person validation provides a practical basis for evaluating that connection in everyday use.

%% file: sections/appendix.tex
\section{Scaffold Selection}
\label{app:scaffolds}

The four compared prompts were direct estimation, COG, a neural scaffold (NEURO), and COMBINED. COG and COMBINED ranked above the alternatives (Table~\ref{tab:prompteval}). Event-structure coding identified 17 single-component and 13 multi-component scenarios; COG was preferred for 14 of the former and COMBINED for 12 of the latter.

\begin{table}[ht]
\centering\small
\begin{tabular}{@{}lrrrrr@{}}
\toprule
Prompt & Rank 1 & Rank 2 & Rank 3 & Rank 4 & Mean rank \\
\midrule
COG & 39 & 26 & 19 & 6 & 1.91 \\
COMBINED & 34 & 31 & 8 & 17 & 2.09 \\
NEURO & 13 & 12 & 51 & 14 & 2.73 \\
Direct estimation & 4 & 21 & 12 & 53 & 3.27 \\
\bottomrule
\end{tabular}
\caption{Overall rankings across 90 scenario--rater combinations.}
\label{tab:prompteval}
\end{table}

\section{Rolling Memory and Input Processing}
\label{app:memory}
\label{app:input}

The 150-word memory block retains up to three recurring themes, the two most recent user corrections, adoption and outcome information for attempted regulation techniques, and one-line summaries of the previous four days. The block enters both inference and conversation calls.

Spoken input is transcribed through the browser's Web Speech API and can be edited before submission. The study system stores transcripts rather than raw audio. Inference runs concurrently with the SAM interface, and the initial output is withheld until the participant submits the pre-guidance ratings. Post-guidance estimation uses the full transcript and runs in the background while the participant completes self-report and feedback.

\section{Regulation Strategy Selection}
\label{app:strategy}

Table~\ref{tab:strategy} crosses intensity and controllability to give the primary prior for each strategy family. Temporal phase can override these priors. Anticipatory events favor situational action, while aftermath and evening reflection can favor cognitive change or response modulation. Situation selection is omitted because the user has already chosen the event to recount.

\begin{table}[ht]
\centering\small
\begin{tabular}{@{}lll@{}}
\toprule
 & High controllability & Low controllability \\
\midrule
Low intensity & Situation modification & Cognitive change \\
High intensity & Attentional deployment & Response modulation \\
\bottomrule
\end{tabular}
\caption{Primary strategy-selection priors, refined by event phase and context.}
\label{tab:strategy}
\end{table}

\section{Study and Per-Session Protocol}
\label{app:protocol}

\begin{figure}[ht]
\centering
\includegraphics[width=\linewidth]{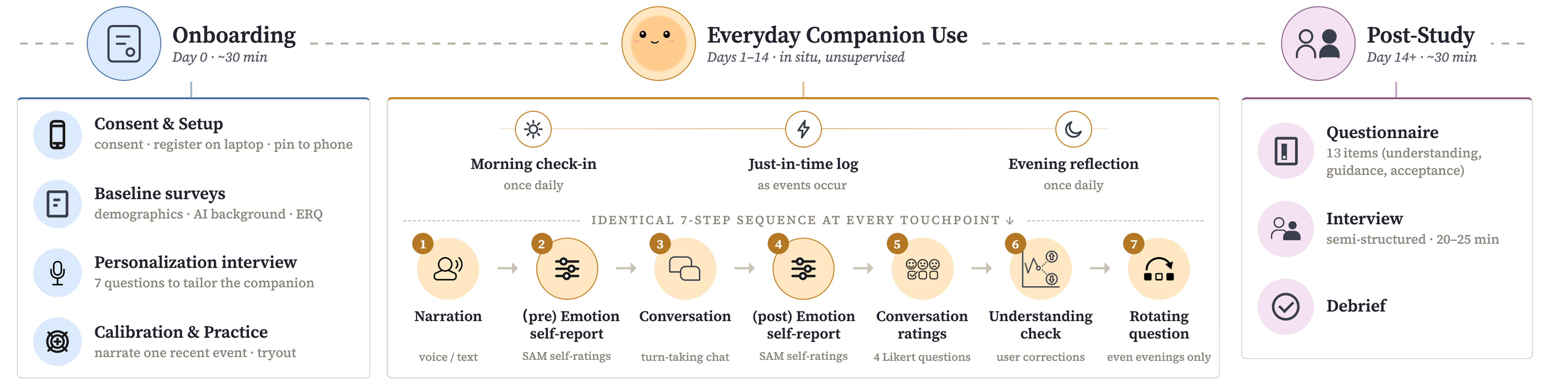}
\caption{Study procedure from onboarding through daily use and the post-study session. Each event pairs initial and post-guidance estimates with corresponding SAM self-reports.}
\label{fig:procedure}
\end{figure}

Figure~\ref{fig:procedure} summarizes the study procedure.

\section{Analysis Populations and Statistical Models}
\label{app:analysis}

\paragraph{Analysis populations.}
Each session carries initial and post-guidance estimates, paired SAM reports, and an understanding rating. Frequency models describe active days; days without a session were not zero-filled. Session-level descriptive statistics and participant-level inferential summaries weight observations differently.

\paragraph{Estimator comparison.}
For the NRC-VAD baseline, matched English word scores are averaged within each narration and mapped to the 1--9 SAM scale by linear regression using five randomly assigned folds; held-out predictions supply the lexicon metrics.

\paragraph{Repeated-measures inference.}
Wilcoxon tests are two-sided, with exact probabilities enumerating signs of nonzero differences; Holm-adjusted $p$ values are $1.53\times10^{-5}$, $.000404$, and $1.14\times10^{-5}$ for valence, arousal, and dominance. Strategy comparisons first average valence change within each participant and strategy, then apply a Friedman test with 100{,}000 within-participant permutations. Correction comparisons pair corrected and uncorrected mean changes for the 12 participants with both.

\begin{table}[ht]
\centering\small
\begin{tabular}{@{}lrrrrr@{}}
\toprule
Contrast & Sessions & Mean change & $SD$ & $W$ & $p_{\mathrm{Holm}}$ \\
\midrule
$\Delta$ Valence & 1093 & $+.525$ & 1.243 & 1 & $<.001$ \\
$\Delta$ Arousal & 1093 & $+.334$ & 1.393 & 11 & $<.001$ \\
$\Delta$ Dominance & 1093 & $+.469$ & 1.132 & 0 & $<.001$ \\
\midrule
$\Delta V \mid V_{\mathrm{before}}\leq4$ & 314 & $+1.344$ & 1.455 & --- & --- \\
$\Delta A \mid A_{\mathrm{before}}\geq7$ & 218 & $-.427$ & 1.243 & --- & --- \\
$\Delta V \mid V_{\mathrm{before}}\geq7$ & 418 & $-.108$ & .857 & --- & --- \\
\bottomrule
\end{tabular}
\caption{Within-session SAM change (after minus before). Means and SDs describe sessions; conditional rows are descriptive.}
\label{tab:delta}
\end{table}

Correlations retain pooled session coefficients and use 10{,}000 whole-participant bootstrap resamples for pointwise 95\% intervals (seed 20260906). Understanding--error Spearman intervals are $[-.165,.150]$ for valence, $[-.076,.134]$ for arousal, and $[-.052,.256]$ for dominance. The understanding--valence-change Pearson interval is $[.061,.208]$.

\paragraph{Longitudinal models.}
Session outcomes use random intercepts for participant and participant-day. Sessions per active day use participant-day aggregates and a participant random intercept. Counts use negative-binomial models with a log link; narration characters and user characters per turn use Gaussian models of log length; absolute valence error uses a Gaussian model of log(1+error). The five subjective ratings use Gaussian models on their observed 1--7 scales. Benjamini--Hochberg correction covers the ten tests separately within the day and week families. Daily percentages for log-length outcomes are calculated as $100(\exp(b)-1)$.

\paragraph{Strategy feedback and qualitative evidence.}
Likelihood-ratio tests compare strategy-feedback models with and without strategy; $\chi^2(3)$ statistics are 3.00, 2.98, 3.72, and 4.33 for helpfulness, feasibility, willingness, and warmth. Interview theme counts refer to distinct participants and can overlap. Adoption counts describe 44 probe responses.

\begin{table}[ht]
\centering\small
\begin{tabular}{@{}lrrrrrr@{}}
\toprule
Selected strategy & $n$ & $\Delta V$ & Helpful & Feasible & Willing & Warm \\
\midrule
Situation modification & 317 & $+.672$ & 5.03 & 5.28 & 5.24 & 5.09 \\
Cognitive change & 246 & $+.585$ & 5.16 & 5.18 & 5.09 & 5.29 \\
Response modulation & 256 & $+.453$ & 5.01 & 5.20 & 5.09 & 5.13 \\
Attentional deployment & 274 & $+.369$ & 5.00 & 5.04 & 5.10 & 5.09 \\
\bottomrule
\end{tabular}
\caption{Unadjusted session means by selected strategy. Feedback uses 1--7 scales. Strategy assignment followed context.}
\label{tab:strategy-feedback}
\end{table}

\section{Coarse-Grained Emotion Agreement}
\label{app:coarse}

\begin{table}[ht]
\centering\small
\begin{tabular}{@{}lrrr@{}}
\toprule
Dimension & Accuracy & Chance & $\kappa$ \\
\midrule
Valence & .780 & .524 & .539 \\
Arousal & .620 & .551 & .154 \\
Dominance & .715 & .501 & .430 \\
\bottomrule
\end{tabular}
\caption{Binary agreement (1--5 / 6--9) for 1{,}093 initial estimate--self-report pairs. Chance agreement is computed from class marginals.}
\label{tab:coarse}
\end{table}

Arousal cut points at 3, 4, 5, 6, and 7 produced $\kappa$ values of .105, .201, .154, .086, and .037, respectively.